\documentclass[10pt]{article}

\usepackage{amsmath} 
\usepackage{amssymb,latexsym} 

\title{\large Nonperturbative Newtonian coupling in Einstein gravity with Dirac fields}
\author{L. N. Granda\thanks{e-mail: ngranda@univalle.edu.co} \\ {\it Departamento de F\'isica, Universidad del Valle, A.A. 25360},\\ {\it Cali, Colombia}}

\begin{document}
\maketitle
{\renewcommand{\abstractname}{}\begin{abstract}
\textbf{Summary. --} We investigate the exact renormalization group (RG) in Einstein gravity coupled to N-component spinor field, working in the effective average action formalism and background field method. The truncated evolution equation is obtained for the Newtonian and cosmological constants. We have shown that screening or antiscreening behaviour of the gravitational coupling depends crucially on the number of spinor components.
\\
\hspace{-0.5cm}{\small PACS. 04.62-v - Quantum field theory in curved spacetime}\\
{\small PACS. 04.60-m- Quantum gravity.}\\ 
{\small PACS. 12.10Dm- Unified theories and models of strong electroweak interactions.}
\vspace{1.0cm}

\end{abstract}

There has recently been much activity in the study of nonperturvative RG dynamics in field theory models ( for recent review, see \cite{BONINI} ). One of the versions of nonpertubative RG based on the effective average action has been developed in ref.\cite{REUTER} for Einstein gravity ( the gauge dependence problem in this formalism has been studied in ref. \cite{FALKENBERG} ) and in \cite{GRANDA} for gauged supergravity. The nonperturbative RG equation for cosmological and Newtonian coupling constants have been obtained in ref.\cite{REUTER}  for the Einstein gravity and in ref \cite{BYTSENKO1} for $R^2$ gravity. The comparison between quantum correction to Newtonian coupling from nonperturvative RG \cite{REUTER} and from effective field theory technique \cite{DONOGHUE} or  $R^{2}$-gravity \cite{ELIZALDE1} has been done.\\
It is quite interesting to study nonperturbative RG (or evolution equation) for the gravity in the presence of $N-$component spinor field in order to evaluate the influence of the number of spinors in the nonperturvative behaviour of the Newtonian and cosmological constants. We follow the formalism of ref. \cite{REUTER} developed for gravitational theories. The basic elements of it are background field method ( see \cite{BUCHBINDER} for a review ) and the truncated nonperturbate evolution equation for the effective average action \cite{REUTER}. We will start from a theory with the following action:\\

\begin{equation}
S=\int d^{4}x \sqrt{g}\Bigl[\frac{1}{16\pi \bar G}\bigr(-R+2\lambda  \bigl)+\bar \psi^{i} i\gamma ^{\mu }\nabla _{\mu } \psi_{i}  \Bigr],
\end{equation}

where $i=,...,N$ and nonrenormalizable Einstein gravity is considered to be valid below some UV scale $\Lambda $. We will use the truncation \cite{REUTER, FALKENBERG}
\begin{equation*}
\bar G\rightarrow G_{k}=\frac{\bar G}{Z_{Nk}},  \hspace{0.5cm}\lambda \rightarrow \lambda _{k}
\end{equation*}

Following the approach of ref. \cite{REUTER} we will write the evolution equation for the effective average action  $\Gamma _{k}[g,\bar g]$ defined at nonzero momentum ultraviolet $k$ below some cut-off $\Lambda_{cut-off}$. The truncated form of such evolution equation has the form

\begin{equation}\label{gran2}
\begin{aligned}
\partial_{t}\Gamma _{k}[g,\bar g]=&\frac{1}{2}Tr\Bigl[(\Gamma^{(2)} _{k}[g,\bar g]+R^{grav}_{k}[\bar g])^{-1}\partial_{t}R^{grav}_{k}[\bar g]\Bigl] \\ &-Tr\Bigl[(-M[g,\bar g]+R^{gh}_{k}[\bar g])\partial_{t} R^{gh}_{k}[\bar g]\Bigr],
\end{aligned}
\end{equation}

where $t=Ln k, k$ is the nonzero momentum scala, $R_{k}$ are cutt-off, $M[g,\bar g]$ are ghost operators, $\bar g_{\mu \nu}$ is the background metric, $g_{\mu \nu }=\bar g_{\mu \nu }+h_{\mu \nu }$  where $h_{\mu \nu }$ is the quantum field. $\Gamma ^{(2)}_{k}$ is the hessian of $\Gamma _{k}[g,\bar g]$ with respect to $g_{\mu \nu }$ at fixed $\bar g_{\mu \nu }$ (for more details, see \cite{REUTER}).\\
Projecting the evolution equation on the space with low-derivatives terms, one gets the left-hand side of the truncated evolution equation (\ref {gran2}) as follows:
\begin{equation}\label{gran3}
\partial_{t}\Gamma _{k}[g,\bar g]=2k^{2}\int d^{4}x\sqrt{g}[-R(g)\partial_{t}Z_{Nk}+2\partial_{t}(Z_{Nk}\lambda _{k})],
\end{equation}
with $k^{2}=1/32\pi \bar G$. The initial conditions for $Z_{Nk}$, $\lambda _{k}$ are chosen in  the same way as in $\cite{REUTER}$.\\
On the right-hand side of the evolution equation $(2)$ we need the second functional derivate of $\Gamma _{k}[g,\bar g]$ at fixed background $\bar g_{\mu \nu }$. A useful choice is De Sitter space where the curvature $R$ is the external parameter characterizing the space. In such a space we have for the gravitational sector of eq. (1)

\begin{equation}\label{gran4}
\begin{aligned}
\Gamma ^{(2)grav}_{k}[g, g]&=Z_{Nk}k^{2}\int d^4x\sqrt{g} \Bigl\{\frac{1}{2}\hat h_{\mu \nu }\bigr[-\Box-2\bar \lambda _{k}+\frac{2}{3}R\bigr]\hat h_{\mu \nu }-\frac{1}{8}\phi [-\Box-2\bar \lambda _{k}]\phi\Bigl\}.
\end{aligned}
\end{equation}
The complete average effective action with $\bar g=g$ containing gravitational, ghosts and spinor contributions is calculated with the final result

\begin{equation}
\begin{aligned}
\Gamma_{k}&=\frac{1}{2}Tr_{\textit{T}}Ln \Biggl[Z_{Nk}\Big(-\Box -2\lambda _{k}+\frac{2}{3}R+k^{2}R^{(0)}(-\Box /k^{2})\Bigr)\Biggr]\\ &+\frac{1}{2}Tr_{S}Ln\biggl[Z_{Nk}Ln\Bigl(-\Box -2\lambda _{k}+k^{2}R^{(0)}(-\Box /k^{2})\Bigr)\Biggr]\\ &-Tr_{V}Ln\Biggl[\Bigl(-\Box -\frac{1}{4}R+k^{2}R^{(0)}(-\Box /k^{2})\Bigr)\Biggr]\\ &-\frac{N}{2}Tr_{1/2}Ln\Biggl[Z_{Nk}\Biggl(-\Box +\frac{1}{4}R+k^{2}R^{(0)}(-\Box/k^{2})\Biggr)\Biggr].
\end{aligned}
\end{equation}

Note that the above calculation may be easily extended to hyperbolic background where the main part of the effective action is well known \cite{BYTSENKO}. Now we want to find the RHS of the evolution equation. To this end, we differentiate the average action (5) with respect to $t$. Then we expand the operators in (5) with respect to the curvature $R$ because we are noly interested in terms of orden $\int d^{4}x\sqrt{g}$ and $\int d^{4}x\sqrt{g}R$:
\begin{equation}
\begin{aligned}
\partial_{t}\Gamma _{k}[g,g]=& Tr_{T}\biggl[{\cal N}\Bigl( A+\frac{2}{3}R\Bigr)^{-1} \Biggr]+Tr_{S}[{\cal N}A^{-1}]\\ & -2Tr_{V}\Biggl[ {\cal N}_{0}\Bigl( {\cal A}_{0}-\frac{1}{4}R     \Bigr)^{-1} \Biggr]+N Tr_{S}\Bigl[{\cal N}_{0}({\cal A}_{0}+\frac{1}{4}R)^{-1}\Bigr]
\end{aligned}
\end{equation}
where\\
\begin{equation*}
{\cal N}=\frac{\partial_{t}[Z_{Nk}k^{2}R^{(0)}(z)]}{Z_{Nk}},
\end{equation*}

\begin{equation}
{\cal A}=-\Box +k^{2}R^{(0)}(z)]-2\lambda _{k}.
\end{equation}

The operators ${\cal N}_{0}$ and ${\cal A}_{0}$ are defined from (7) with $\lambda _{k}=0$ and $Z_{Nk}=1$ (see \cite{REUTER, FALKENBERG}). Here the variable $z$ replaces $-\Box/k^{2}$ and $\eta _{N}(k)=-\partial_{t}(Ln Z_{Nk})$. Note that as a cut-off we use the same function as in ref.\cite{REUTER}: $R^{(0)}(z)=\frac{z}{exp[z]-1}$. The above steps lead then to
\begin{equation}
\begin{aligned}
\partial_{t}\Gamma _{k}[g,g]=& Tr_{T}\biggl[{\cal N}\Bigl(A\Bigr)^{-1} \Biggr]+Tr_{S}[{\cal N}A^{-1}] -2Tr_{V}\Biggl[ {\cal N}_{0}{\cal A}_{0}^{-1} \Biggr]\\ &-N Tr_{1/2}[{\cal N}_{0}{\cal A}^{-1}_{0}]-R \Bigl\{ \frac{2}{3}Tr_{T}[{\cal N}{\cal A}^{-2}]+\frac{1}{2}Tr_{V}[{\cal N}_{0}{\cal A}_{0}^{-2}]\\ &-\frac{1}{4} N Tr_{1/2}[{\cal N}_{0}{\cal A}_{0}^{-2}]\Bigr\}+O(R^{2}),
\end{aligned}
\end{equation}
As a next step we evaluate the traces. We use the heat kernel expansion which for an arbitrary function of the covariant Laplacian $W(D^{2})$ reads
\begin{equation}
\begin{aligned}
Tr_{j}[W(-D^{2})]=& (4\pi )^{-2}tr_{j}(I)\Biggl\{ Q_{2}[W]\int d^{4}x\sqrt{g}\\ & + \frac{1}{6}Q_{1}[W]\int d^{4}x\sqrt{g}R+O(R^{2})\Biggr \},
\end{aligned}
\end{equation}
where by $I$ we denote the unit matrix in the space of field on which $D^{2}$ acts. Therefore $tr_{j}(I)$ simply counts the number of independent degrees freedom of the field. The sort $j$ of fields enters $(16)$ via $tr_{j}(I)$ only. Therefore, we will drop the index $j$ after the evaluation of the traces in the heat kernel expansion. Note that one could use also zeta-regularization in the above expression [10].\\
The functional $Q_{n}$ are the Mellin transforms of W, 
\begin{equation}
\begin{aligned}
Q_{n}[W]=\frac{1}{\Gamma(n)}\int^{\infty}_{0} dzz^{n-1}W(z) \hspace{1cm}(n > 0).
\end{aligned}
\end{equation}
Now we have to perform the heat kernel expansion (9) in eq.(8). This leads to a polynomial in $R$, which is the RHS of the evolution equation (2).\\
By the comparison of coefficients with the LHS of the evolution equation (3), we obtain  at the order $\int d^{4}x\sqrt{g}$
\begin{equation}
\begin{aligned}
\partial_{t}(Z_{Nk}\bar  \lambda _{k})=\frac{1}{4k^{2}}\frac{1}{(4\pi )^{2}}\{ 10Q_{2}[{\cal N}/{\cal A}]-8Q_{2}[{\cal N}_{0}/{\cal A}_{0}]-4NQ_{2}[{\cal N}_{0}/{\cal A}_{0}]\}
\end{aligned}
\end{equation}
and at the order $\int d^{4}x\sqrt{g}R.$
\begin{equation}
\begin{aligned}
\partial_{t}Z_{Nk}&=-\frac{1}{12k^{2}}\frac{1}{(4\pi )^{2}}\{10Q_{1}[{\cal N}/{\cal A}]-8Q_{1}[{\cal N}_{0}/{\cal A}_{0}]\\ &-4NQ_{1}[{\cal N}_{0}/{\cal A}_{0}]-36Q_{2}[{\cal N}/{\cal A}^{2}]-12Q_{2}[{\cal N}_{0}/{\cal A}^{2}_{0}]+6 NQ_{2}[{\cal N}_{0}/{\cal A}_{0}^{2}]\}.
\end{aligned}
\end{equation}

The cutt-off-dependent integrals are defined in \cite{REUTER, FALKENBERG}

\begin{equation*}
\varPhi ^{P}_{n}(w)=\frac{1}{\Gamma (n)}\int ^{\infty }_{0}dzz^{n-1}\frac{R^{(0)}(z)-zR^{(0)'}(z)}{[z+R^{(0)}(z)+w]^{P}},
\end{equation*}

\begin{equation}
\tilde \varPhi ^{P}_{n}(w)=\frac{1}{\Gamma (n)}\int ^{\infty }_{0}dzz^{n-1}\frac{R^{(0)}(z)}{[z+R^{(0)}(z)+w]^{P}},
\end{equation}
for $n > 0$. It follows that $\varPhi^{P}_{0}(w)=\tilde \varPhi^{P}_{0}(w)=(1+w)^{-P} $ for $n=0$. We can rewrite eqs. (11) and (12) in terms of $\varPhi $ and $\tilde  \varPhi $. This leads to the following system of equations:
\begin{equation}
\begin{aligned}
\partial_{t}(Z_{Nk}\bar \lambda _{k})&=\frac{1}{4k^{2}}\frac{1}{(4\pi )^{2}}k^{4}\{10Q^{1}_{2}(-2\bar \lambda _{k}/k^{2})-8\varPhi^{1}_{2}(0)\\ & -4N\varPhi^{1}_{2}(0)-5\eta _{N}(k)\tilde \varPhi^{1}_{2}(-2\bar \lambda _{k}/k^{2})\},
\end{aligned}
\end{equation}

\begin{equation}
\begin{aligned}
\partial_{t}Z_{Nk}&=-\frac{1}{12k^{2}}\frac{1}{(4\pi )^{2}}k^{2}\{10\varPhi^{1}_{1}(-2\bar \lambda _{k}/k^{2})-(4N+8)\varPhi^{1}_{1}(0)\\ &-36\varPhi^{2}_{2}(-2\bar \lambda _{k}/k^{2})-(6 N-12)\varPhi^{2}_{2}(0)\\ &-5\eta _{N}(k)\tilde \varPhi^{1}_{1}(-2\bar \lambda _{k}/k^{2})+18\eta _{N}(k)\tilde \varPhi^{2}_{2}(-2\bar \lambda _{k}/k^{2})\} .
\end{aligned}
\end{equation}

Now we introduce the dimensionless, renormalized Newtonian constant and cosmological constant

\begin{equation}
g_{k}=k^{2}G_{k}=k^{2}Z^{-1}_{Nk}\bar G,\hspace{1cm}\lambda _{k}=k^{-2}\bar \lambda _{k}.
\end{equation}
Here $G_{k}$ is the renormalized Newtonian constant at scale $k$. The evolution equation for $g_{k}$ reads then
\begin{equation}
\partial_{t} g_{k}=[2+\eta _{N}(k)]g_{k}.
\end{equation}
from $(15)$ we find the anomalous dimension $\eta _{N}(k)$
\begin{equation}
\eta _{N}(k)=g_{k}B_{1}(\lambda _{k})+\eta _{N}(k)g_{k}B_{2}(\lambda _{k})
\end{equation}
where\\
\begin{equation}
\left\{\begin{aligned}&B_{1}(\lambda _{k})=\frac{1}{6\pi }[10\varPhi^{1}_{1}(-2\lambda _{k})-4(N+2)\varPhi^{1}_{1}(0)-36\varPhi^{2}_{2}(-2\lambda _{k})+(6N-12)\varPhi^{2}_{2}(0)],\\ &B_{2}(\lambda _{k})=\frac{1}{6\pi }[18\tilde \varPhi^{2}_{2}(-2\lambda _{k})-5\tilde \varPhi^{1}_{1}(-2\lambda _{k})].
\end{aligned}\right.
\end{equation}\\
Solving (18)\\
\begin{equation}
\eta _{N}(k)=\frac{g_{k}B_{1}(\lambda _{k})}{1-g_{k}B_{2}(\lambda _{k})}
\end{equation}

we see that the anomalous dimension $\eta _{N}$ is a nonperturbative quatity. From $(14)$ we obtain the evolution equation for the cosmological constant
\begin{equation}
\begin{aligned}
\partial _{t}(\lambda _{k})=&-[2-\eta _{N}(k)]\lambda _{k}+\frac{1}{2\pi }g_{k}[10\varPhi^{1}_{2}(-2\lambda _{k})+\\ &-4(N+2)\varPhi^{1}_{2}(0)-5\eta _{N}(k)\tilde \varPhi^{1}_{2}(-2\lambda _{k})].
\end{aligned}
\end{equation}
Equations $(17)$ and $(21)$ together with $(19)$ give the system of differential equations for two $k$-dependent constants $\lambda _{k}$ and $g_{k}$. These equations determine the value of the running Newtonian constant and cosmological constant at the scale $k\ll \Lambda _{cut-off}.$ The  above evolution equations include nonperturbative effects which go beyond  a simple one-loop calculation.\\
Next, we estimate the qualitative behaviour of the running Newtonian constant as the above system of RG equations is too complicated and cannot be solved analytically. To this end we assume that the cosmological constant is much smaller than the IR cut-off scale, $\lambda _{k}\ll k^{2}$, so we can put $\lambda _{k}=0$ that simplify eqs. $(17)$ and $(19)$. After that, we make an expansion in powers of $(\bar Gk^{2})^{-1}$ keeping only the first term (\textit{i,e.} we evaluate the function $\varPhi^{P}_{n}(0)$) and $\tilde \varPhi^{P}_{n}(0)$) and finally obtain (with $g_{k}\sim k^{2}\bar G$)
\begin{equation}
\begin{aligned}
G_{k}=G_{0}[1-w\bar Gk^{2}+...],
\end{aligned}
\end{equation}
where \\
\begin{equation}
\begin{aligned}
w&= \frac{1}{12\pi }[(48-6 N)\varPhi^{2}_{2}(0)+(4N-2)\varPhi^{1}_{1}(0)]\\ &=\frac{1}{12\pi }\Bigl[ (48-6N)+(4N-2)\frac{\pi ^{2}}{6}\Bigr],
\end{aligned}
\end{equation}\\
where $\varPhi^{1}_{1}(0)=\pi ^{2}/6$ and $\varPhi^{2}_{2}(0)=1$.\\
In the case of Einstein gravity, a similar solution has been obtained in refs. \cite{REUTER, FALKENBERG}.\\
From $(18)$ we can see that for any values of $N$, $w$ will be positive, which means that Newton's constant decreases as $k^{2}$ increases; it is the same sing as for non-Abelian gauge coupling in Yang-Mills theory. We see that gravity is antiscreening, being larger at large distances. Hence, we proved that antiscreening behaviour of the gravitational constant is expected in the presence of spinor fields, growing with $N$. That may lead to interesting cosmological consequences in the early Universe.\\\\
\centerline{***}\\\\
I would like to thank \textsl{COLCIENCIAS} for financial support. I am grateful to \textit{S. D. Odintsov} for stimulating discussions.\\

\thispagestyle{empty}

\begin{thebibliography}{*}
\addcontentsline{toc}{chapter}{Bibliografía}
\bibitem{BONINI}B{\small ONINI M}., D'{\small ATTANASIO} M. and M{\small ARCHESINI} G., \textit{Nucl. Phys}. B, \textbf{437}(1995) 163; M{\small ORRIS} T. R., \textit{Phys. Lett. B,}  \textbf{329} (1994) 241; E{\small LLWANGER} U. and V{\small ERGARA} L., \textit{Nucl. Phys. B,} \textbf{398} (1993) 52; R{\small EUTER} M. and W{\small ETTERICH} C., \textit{Nucl. Phys. B,} \textbf{427} (1994) 291; L{\small IAO} S. B., P{\small OLONYI} J. and X{\small U} D.,  \textit{Phys. Rev. D,} \textbf{51} (1995) 748.

\bibitem{REUTER}R{\small EUTER} M., D{\small ESY} \textbf{96-080}, hep-th 9605030. 
\bibitem{FALKENBERG}F{\small ALKENBERG}. S. and O{\small DINTSOV} S. D., hep-th 9612019, I{\small JMPA}, to be published.
\bibitem{DONOGHUE} D{\small ONOGHUE} J., phys. Rev. D, \textbf{50}(1994)3874; H{\small AMBER} H. W. and L{\small IU} S., Phys. Lett. B, \textbf{357}(1995) 51; M{\small UZINICH} I. and V{\small okos} S., Phys. Rev. D, \textbf{52}(1995) 3472; M{\small ODANESE} G., Phys. Lett. B, \textbf{325} (1994) 354.
\bibitem{GRANDA}G{\small RANDA} L. N. and O{\small DINTSOV} S. D., \textit{Phys. Lett. B,} \textbf{409} (1997) 206.
\bibitem{BYTSENKO1}B{\small YTSENKO} A. A., O{\small DINTSOV} S. D. and G{\small RANDA} L. N., J{\small ETP} Lett., \textbf{65} (1997) 600

\bibitem{BUCHBINDER} B{\small UCHBINDER} I. L., O{\small DINTSOV} S. D. and  S{\small HAPIRO} I. L., \textit{Effective Action in Quantum Gravity} (IOP Publishing, Bristol) 1992.



\bibitem{ELIZALDE1}E{\small LIZALDE} E., L{\small OUSTO} C., O{\small DINTSOV} S. D. and R{\small OMEO } A., Phys. Rev. D, \textbf{52} (1995) 2202.

\bibitem{BYTSENKO}B{\small YTSENKO} A. A., O{\small DINTSOV} S. D. and Z{\small ERBINI} S., Phys. Lett. B, \textbf{336} (1994) 355.

\bibitem{ELIZALDE}E{\small LIZALDE} E., O{\small DINTSOV} S. D. and R{\small OMEO } A., B{\small YTSENKO} A. A., Z{\small ERBINI} S., Zeta Regularization tecnique with Aplications (World Sci., Singapore) 1994.


\end{thebibliography}
\end{document}